\documentclass[12pt,twoside]{article}
\usepackage{fleqn,espcrc1}

\usepackage{graphicx}
\usepackage[figuresright]{rotating}
\usepackage{wrapfig}

\newcommand{\AmS}{{\protect\the\textfont2
  A\kern-.1667em\lower.5ex\hbox{M}\kern-.125emS}}

\title{Effective Field Theory and Nuclear Mean-Field Models}

\author{R.~J.\ Furnstahl\address{Department of Physics, Ohio State University, 
        Columbus, OH 43210}%
        \thanks{Supported in part by NSF grants PHY--9511923 and  PHY--9800964.}
        and 
        Brian D. Serot\address{Department of Physics, Indiana University,
        Bloomington, IN 47405}%
		\thanks{Supported in part by the DOE under Contract 
		  No.~DE-FG02-87ER40365.}}
       
\begin{document}

% typeset front matter
\maketitle

\smallskip

\begin{abstract}
The implications of an effective field theory (EFT) interpretation of
nuclear mean-field phenomenology are reviewed.
\end{abstract}

\medskip
\noindent\rule{2.5in}{.5pt}
\bigskip
%\section{OVERVIEW}

Recent work has demonstrated 
that effective field theory (EFT) concepts and methods
can explain the successes and limitations of mean-field
models of nuclear structure and 
reactions \cite{Friar96,Furnstahl96,Serot97,Rusnak97,Furnstahl97}. 
For example, coefficients in successful relativistic (QHD) 
and nonrelativistic (Skyrme) mean-field models are consistent
with naive dimensional analysis (NDA) and naturalness, as expected in
low-energy effective field theories of QCD \cite{Georgi84}.

NDA implies
an expansion of the mean-field energy density
in $\rho/f_\pi^2\Lambda$ with coefficients of order unity 
(``naturalness''), where $\rho$ is the nuclear density,
$f_\pi \approx 93\,$MeV is the pion-decay constant, and $\Lambda$ is the
scale of non-Goldstone boson physics (500 to 1000\,MeV) \cite{Friar96,Serot97}. 
This expansion
parameter is between 1/7 and 1/4 at equilibrium nuclear density.  
Nuclear matter binding energies verify this expansion for 
all models that fit nuclear data accurately.
In addition, 
truncation errors due to omitted terms are predicted.

\begin{figure}[t]
\begin{minipage}[t]{75mm}
\includegraphics[width=70mm]{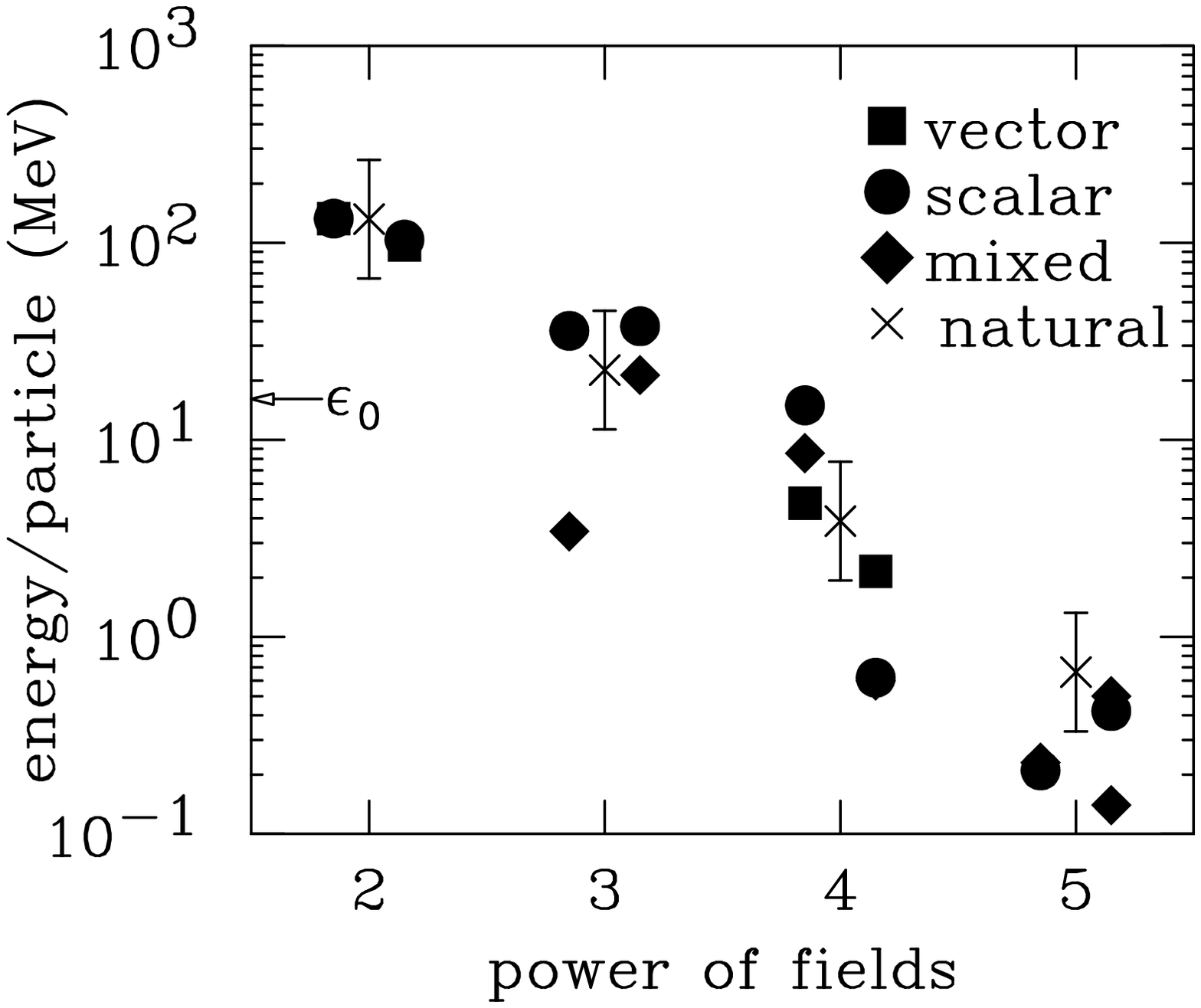}
%\framebox[79mm]{\rule[-26mm]{0mm}{52mm}}
\vspace*{-.3in}
\caption{Energy contributions to two QHD models,
  with estimates based on NDA and natural coefficients
  (error bars).}
\label{fig:vmd}
\end{minipage}
\hspace{\fill}
\begin{minipage}[t]{75mm}
%\framebox[74mm]{\rule[-26mm]{0mm}{52mm}}
\includegraphics[width=68mm]{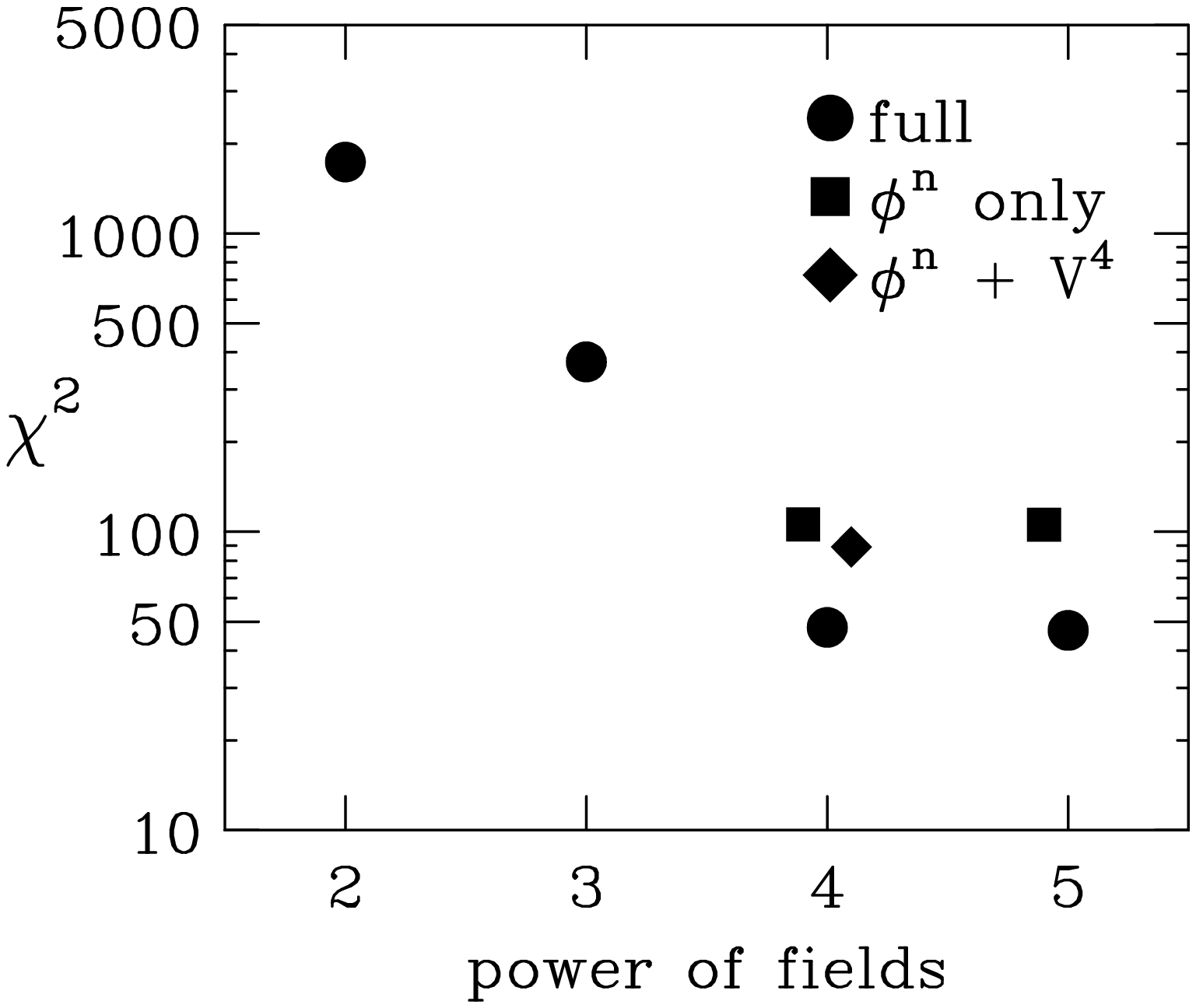}
\vspace*{-.3in}
\caption{$\chi^2$ values for QHD meson models, 
  according to the truncation level.}
\label{fig:chisq}
\end{minipage}
\vspace*{-.25in}
\end{figure}

In Fig.~\ref{fig:vmd}, individual contributions to the energy
per particle from terms in the energy functional of two general QHD
models are plotted against
the net power of the mean meson fields.  
In an EFT, one expects {\it all\/} terms consistent with underlying
symmetries, excluding redundancies that can be removed by redefining the
fields.  The models shown include all terms
in the expansion implied by NDA with coefficients determined by
a $\chi^2$--type optimization to bulk properties of 
magic nuclei \cite{Furnstahl96}.
The crosses are expected values from
NDA estimates, with the error bars spanning a reasonable range of
natural coefficients (from 1/2 to 2).  One model is to the left of
each cross and one is to the right.
The  hierarchy of contributions predicted by NDA is manifest.

In Fig.~\ref{fig:chisq}, the impact of different model truncations
is shown by plotting $\chi^2$ against the maximum power
of fields.
The ``full'' models (which include all terms at a given order) 
show that one needs to go to the fourth power
of fields to get excellent fits, but no further.
The ``$\phi^n$ only'' results, with only scalar fields  for
$n>2$, show that nearly optimal results can
be obtained with just a subset
of terms at each order.  Thus the parameters are underdetermined
by the data. 
These features explain the successes of the most widely used QHD models, 
which add only $\phi^3$ and $\phi^4$ terms to the original Walecka
model.

It is often said that since QHD models use local nucleon fields, they
do not incorporate the effects of
nucleon compositeness.  Indeed, quark-meson coupling (QMC) models
were introduced to remedy this ``deficiency'' \cite{QMC}.
Unlike QHD models, QMC models have density- or field-dependent
couplings, e.g., $g_s\sigma \rightarrow g_s(\sigma)\sigma$.
However, the EFT perspective shows the QMC models are in fact a 
{\it subset\/} of
the general QHD models.
A simple field redefinition,
$g_s(\sigma)\sigma \overline NN \rightarrow g_s\sigma'\overline NN
+ a\sigma'{}^3 + b\sigma'{}^4 + \cdots$, 
moves the QMC vertex corrections into
meson propagator corrections already contained in QHD models.
{\it Observables cannot depend on these off-shell manipulations.\/}

\begin{wrapfigure}{r}{70mm}
\centerline{\includegraphics[width=70mm]{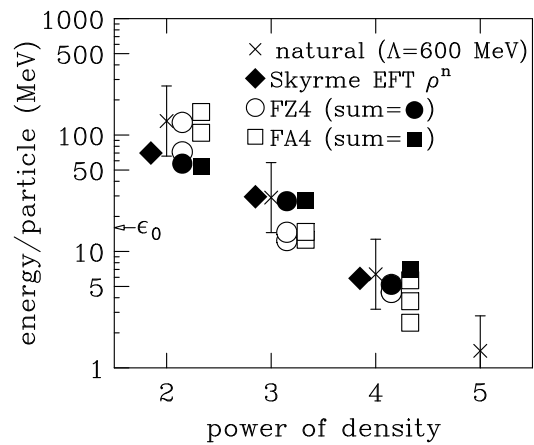}}
\vspace*{-.1in}
\caption{Energy contributions for a generalized Skyrme model
compared to two QHD point-coupling models.}
\label{fig:generalized}
\end{wrapfigure}

In addition, general QHD models incorporate single-nucleon form factors
explicitly in a derivative expansion \cite{Furnstahl96}.
Low energy means low resolution, which implies that a derivative
expansion is efficient.
This is analogous to studying a complicated charge distribution
using long wavelengths.
One  may be sensitive to the dipole moment, and any convenient model
(preferably with a systematic expansion) that reproduces the
moment will work.
However, fitting the dipole moment does not imply that the quadrupole
moment will be correct!  
Similarly, extrapolations of mean-field models to uncalibrated regions
are dangerous.
A virtue of the EFT approach is that the limits of the model
are well understood.

Finally, the EFT interpretation explains how to correctly
account for QCD vacuum physics.  
Naturalness implies that the
numerically important contributions from the vacuum are {\it automatically\/}
incorporated into coefficients already in the model.  
In contrast, models based on a simple $\overline NN$ vacuum (``RHA'')
produce unnatural coefficients \cite{Furnstahl96}.

The ``heavy'' mesons $\omega$, $\rho$, and $\sigma$ that appear in QHD models
have masses at the resolution scale $\Lambda$.  The EFT perspective
implies that we should be able to replace 
these meson interactions with local couplings between nucleons 
(so that the energy density has no heavy meson fields, but only powers of the
scalar and vector densities) with
similar success.
This is indeed the case.
In Fig.~\ref{fig:generalized},
a plot analogous to Fig.~\ref{fig:vmd} is made for two general point-coupling
models (labeled FZ4 and FA4).  Results and conclusions similar
to the QHD meson models are obtained (although one can truncate at third
order and still find a good fit) \cite{Rusnak97}.

\begin{figure}[t]
\begin{minipage}[t]{75mm}
\includegraphics[width=70mm]{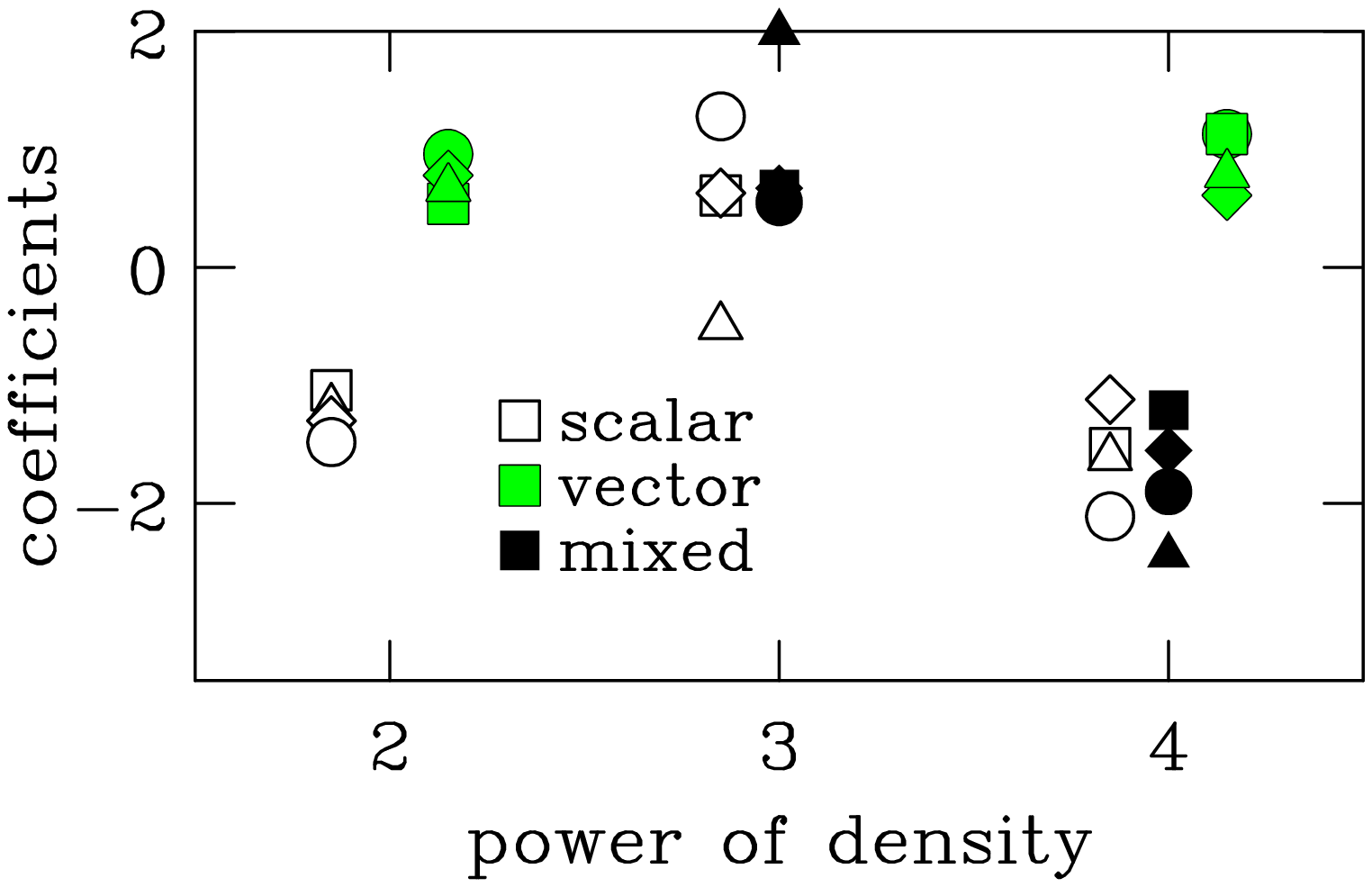}
%\framebox[79mm]{\rule[-26mm]{0mm}{52mm}}
\vspace*{-.3in}
\caption{Coefficients for four good-fit QHD point-coupling models.
Each model is represented by a different shape and the shading shows the
type of term (scalar, vector, or mixed).}
\label{fig:coefficients}
\end{minipage}
\hspace{\fill}
\begin{minipage}[t]{75mm}
%\framebox[74mm]{\rule[-26mm]{0mm}{52mm}}
\includegraphics[width=70mm]{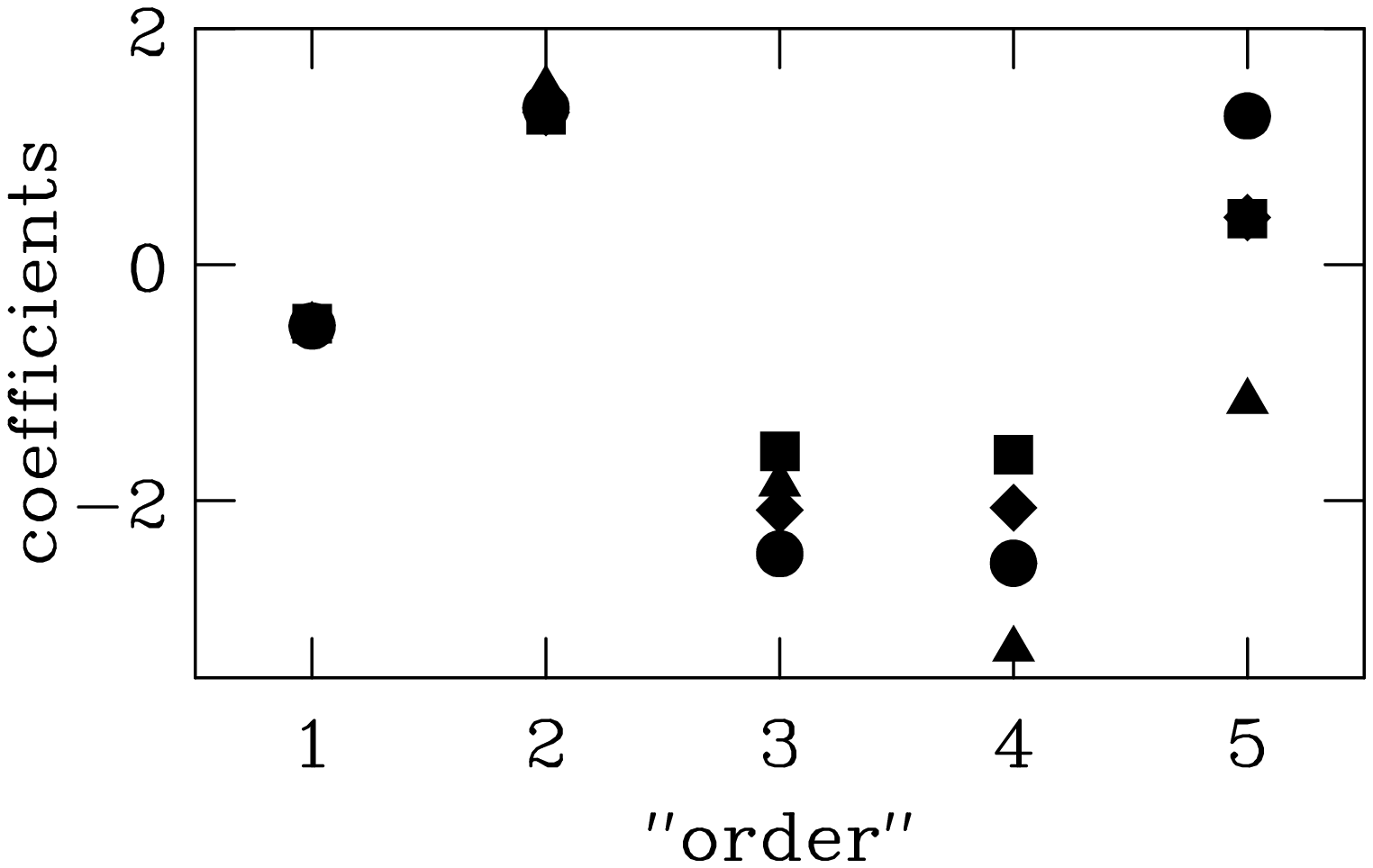}
\vspace*{-.3in}
\caption{Optimal coefficients for the same four models as in 
Fig.~\ref{fig:coefficients}.
The ``order'' is determined by counting powers of $\rho_+$ and
$\rho_-$.}
\label{fig:optimal}
\end{minipage}
\vspace*{-.25in}
\end{figure}

\newcommand{\rhos}{\rho_s}
\newcommand{\rhoB}{\rho_{\scriptscriptstyle\rm B}}

As in Fig.~\ref{fig:vmd}, we see
more than one QHD model with a good fit to nuclei, 
which implies the coefficients
are underdetermined.
This is made explicit in Fig.~\ref{fig:coefficients}, where the coefficients 
from four  models
show large variations even at leading order.
% and vary more at second
%order than at third order.
(Note, however, that all coefficients are natural, i.e., order unity.) 
Can we find a more systematic power counting scheme?
The similar size of the scalar density $\rhos$ and
the vector density $\rhoB$ suggests that we
count instead powers of $\rho_+ \equiv (\rhos+\rhoB)/2$ and
$\rho_- \equiv (\rhos-\rhoB)/2$.
The variations of the new ``optimal'' coefficients
for the four models are shown in Fig.~\ref{fig:optimal}.
Leading orders are very well determined, with a 
systematic increase in uncertainty
until even the sign is undetermined at the highest order shown. 

These results suggest that a nonrelativistic point-coupling EFT,
with an expansion in $\rho \equiv \rhoB$ only,
should work well for bulk nuclear observables.
Indeed, the phenomenologically successful Skyrme
models are of this type.  
Applying NDA shows that they are natural \cite{Furnstahl97}.
Since conventional Skyrme models are truncated at $\rho^3$, a generalized
version with up to $\rho^4$ was fit to nuclei.
The results in Fig.~\ref{fig:generalized} show the same pattern
predicted by NDA, except that the leading term is rather low.
This is explained by comparing to the net results (filled symbols) 
from scalar and
vector ``two-body'' contributions in the relativistic model:
The underlying large mass scale is hidden by cancellations at order $\rho^2$.
Note that higher orders follow the NDA predictions.
A scale of $\Lambda=600\,$MeV is consistent with the trends
in all of the relativistic and nonrelativistic models. 

\def\lsim{\lower0.6ex\vbox{\hbox{$\ \buildrel{\textstyle <}
         \over{\sim}\ $}}}

What about the role of short-range correlations?
Since we fit to observed nuclear properties, 
we {\it do\/} include correlation effects (approximately)
in the
coefficients of the model.
An underlying assumption of our application of NDA to mean-field models
is that the sizes of coefficients are dominated by the short-distance
scales $(r\lsim \Lambda^{-1})$
and not potentially longer-ranged many-body scales.
% $(r \lsim k^{-1}_{\mathrm Fermi})$.  
 Or, at least,
 that the
important many-body and short-distance scales are of similar magnitude.
Density functional theory may provide a framework for 
the systematic
inclusion of correlation effects \cite{DFT}.
Our mean-field models are analogs of the Kohn--Sham formalism, with
local meson fields playing the role of (relativistic) Kohn--Sham potentials.
Correlations are included exactly if the correct functional is used.
Mean-field models approximate this functional with powers of fields
or densities.
The mean-field functional misses possible
nonanalytic terms; a combination of EFT and density functional 
theory may show us how to systematically include them.

In summary, the application of EFT concepts and methods to mean-field
models of nuclei reveals that:
\begin{itemize}
  \item NDA provides an organizational principle at the mean-field level.
  The EFT power counting and the underdetermination of parameters by nuclei
  explain the success of conventional mean-field models \cite{Serot97}.
  \item Mean-field models are approximate implementations of Kohn--Sham
  density functional theory, which means that correlation effects
  are included in simple Hartree calculations.
  \item Vacuum effects, chiral symmetry, and nucleon substructure
  {\it are\/} all included 
  in general QHD models.  This implies that the success of a particular 
  non-QHD model 
  is not necessarily evidence of the reality of the model dynamics.
  \item Ground-state nuclear properties provide information at
  low resolution.
  Models with different degrees of freedom (e.g., four- vs.\ two-component
  nucleons) are simply different organizations of the EFT.  All are
  consistent with NDA. 
\end{itemize}

Work is in progress
to develop a more complete and consistent
power counting, to connect to recent EFT studies of few-nucleon
systems, and to develop density functional theory in an EFT framework.
Explicit pion-loop corrections are being studied along with tests
of general mean-field models
for other observables, such as collective excitations, nuclear currents, and
the isovector response.

\clearpage  % flush all figures!

\end{document}